\documentstyle[sprocl]{article}

\input{psfig}
\def\NPB{{\em Nucl. Phys.} B}
\def\PLB{{\em Phys. Lett.}  B}
\def\PRL{\em Phys. Rev. Lett.}
\def\PRD{{\em Phys. Rev.} D}

\def\be{\begin{equation}}
\def\nn{\noindent}
\def\ie{{\it i.e.}}
\def\eg{{\it e.g.}}

\def\etal{{\it et al.}}
\def\ee{\end{equation}}
\def\bea{\begin{eqnarray}}
\def\eea{\end{eqnarray}}
\begin{document}

\rightline{\vbox{\halign{&#\hfil\cr
SLAC-PUB-8192\cr
July 1999\cr}}}
\vspace{0.8in}

\title{{POLARIZATION ASYMMETRIES IN $\gamma e$ COLLISIONS AND TRIPLE GAUGE 
BOSON COUPLINGS REVISITED}
\footnote{To appear in the {\it Proceedings of the World-Wide Study of 
Physics and Detectors for Future Linear Colliders(LCWS99)}, Sitges, 
Barcelona, Spain, 28 April-5 May 1999}
}

\author{ {T.G. RIZZO}
\footnote{Work supported by the Department of Energy, 
Contract DE-AC03-76SF00515}
}

\address{Stanford Linear Accelerator Center,\\
Stanford University, Stanford, CA 94309, USA}


\maketitle\abstracts{The capability of the NLC run in the $\gamma e$ collision 
mode to probe the CP-conserving $\gamma WW$ and $\gamma ZZ$ anomalous 
couplings through the use of the polarization asymmetry is reviewed. When 
combined with other measurements, very strong constraints on both varieties 
of anomalous couplings can be obtained. We show that these bounds are 
complementary to those that can be extracted from data taken at the LHC.}

In addition to unravelling the source of symmetry 
breaking, one of the most crucial remaining set of tests of the 
structure of the SM will occur at future colliders when precision measurements 
of the various triple gauge boson vertices(TGVs) become 
available~{\cite {rev}}. If new physics arises at or near the TeV scale, 
then on rather general grounds one expects that the deviation of the 
TGVs from their canonical SM values, \ie, the anomalous 
couplings, to be {\it at most} ${\cal O}(10^{-3}-10^{-2})$ with the smaller 
end of this range of values being the most likely. To get to 
this level of precision and beyond, for all of the TGVs, a number of 
different yet complementary reactions 
need to be studied using as wide a variety of observables as possible. 

In the present review we concentrate on the 
CP-conserving $\gamma WW$ and $\gamma ZZ$ anomalous couplings that can be 
probed in the reactions $\gamma e \to W\nu ,Ze$ at the NLC using polarized 
electrons and polarized backscattered laser photons~{\cite {old}}. In the 
$\gamma WW$ case, the anomalous 
couplings modify the magnitude and structure of the already existing SM tree 
level vertex. No corresponding tree level $\gamma ZZ$ vertex exists in 
the SM, although it will appear at the one-loop level. One immediate 
advantage of the $\gamma e\to W\nu$ process over, \eg, 
$e^+e^-\to W^+W^-$ is that the $\gamma WW$ vertex can be trivially isolated 
from the corresponding ones for the $ZWW$ vertex, thus allowing us to probe 
this particular vertex in a model-independent fashion and for the case of 
on-shell photons. We recall that the $CP-$conserving $\gamma WW$ and 
$\gamma ZZ$ anomalous couplings are 
denoted by $\Delta \kappa$, $\lambda$  and $h_{3,4}^0$, 
respectively~{\cite {rev}}.

The use of both polarized electron and photon beams available at the NLC 
allows one to construct a polarization asymmetry, $A_{pol}$, 
which provides a new handle on possibly anomalous TGVs of both the 
$W$ and $Z$. In general the $\gamma e \to W\nu ,Ze$ 
(differential or total) cross sections can be written schematically 
as $\sigma=(1+A_0P)\sigma_{un}+\xi(P+A_0)\sigma_{pol}$, 
where $P$ is the electron's polarization(which we take to be $>0$ for 
left-handed beam polarization), 
$-1\leq \xi \leq 1$ is the Stoke's parameter for the circularly polarized 
photon, and $A_0$ describes the electron's coupling to the relevant gauge 
boson[$A_0=2va/(v^2+a^2)=1$ for $W$'s and $\simeq 0.149$ for $Z$'s]. 
$\sigma_{pol}(\sigma_{un})$ 
represents the polarization (in)dependent contribution to the cross section, 
both of which are functions of only a single dimensionless variable 
after angular integration, \ie, $x=y^2=s_{\gamma e}/M_{W,Z}^2$,  
where $\sqrt {s_{\gamma e}}$ is the $\gamma -e$ center of mass energy. 
Taking the ratio of the $\xi$-dependent to $\xi$-independent terms in the 
expression for $\sigma$ defines the asymmetry $A_{pol}$.

One reason to expect that $A_{pol}$, or $\sigma_{pol}$ itself,  
might be sensitive to modifications in the TGVs due to the presence of the 
anomalous couplings is the Drell-Hearn Gerasimov(DHG) Sum Rule~{\cite {dhg}}.  
In its $\gamma e \to W \nu, Ze$ manifestation, the DHG sum rule implies that
\begin{equation}
\int_{1}^{\infty} {\sigma_{pol}(x)\over {x}} dx = 0 \,,
\end{equation}
for the tree level SM cross section when the couplings of all the 
particles involved in the process are renormalizable and gauge invariant. 
That this integral is zero results from ($i$) the fact that 
$\sigma_{pol}$ is well 
behaved at large $x$ and ($ii$) a delicate cancellation occurs 
between the two 
regions where the integrand takes on opposite signs. This observation is 
directly correlated with the existence of a unique 
value of $x$, \ie, $x_0$,  where $A_{pol}$ vanishes.
For the $W(Z)$ case this asymmetry `zero' occurs at approximately 
$\sqrt {s_{\gamma e}}\simeq 254(150)$ GeV, both 
of which correspond to energies which are easily accessible at the NLC. In the 
$Z$ boson case, unlike that for $W$'s, the SM position of the zero can 
be obtained analytically as a 
function of the cut on the angle of the outgoing electron. 

As discussed in detail in Ref.~{\cite {old}}, the inclusion of anomalous 
couplings not only moves the position of the zero but also forces the 
integral to become non-vanishing and, in most cases, 
{\it logarithmically divergent}. In fact, 
the integral is only finite when $\Delta \kappa+\lambda=0$, the same condition 
necessary for the existence of the radiation amplitude zero~{\cite {raz}}. 
It is 
interesting that the anomalous couplings do not induce additional zeros or 
extinguish the zero completely. 
Unfortunately, since we cannot go to infinite energies we cannot test the DHG 
Sum Rule directly but we {\it are} left with the position of the zero, or more 
generally, the asymmetry itself as a probe of TGVs. In the 
$W$ case the zero position, $x_0$, is found to be far more sensitive to 
modifications in the TGVs than is the zero position in in the $Z$ case. 

\nn
\vspace*{0.1mm}
\hspace*{-0.5cm}
\begin{figure}[htbp]
\centerline{\psfig{figure=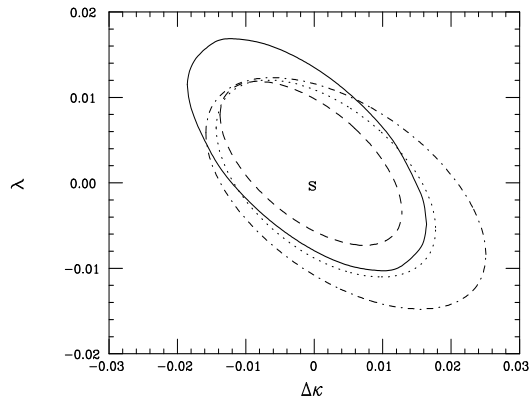,height=6.5cm,width=8cm,angle=90}}
\vspace*{-1.0cm}
\caption{95 $\%$ CL bounds on the $W$ anomalous couplings from the 
polarization asymmetry. The 
solid(dashed, dash-dotted) curves are for a 500 GeV NLC 
assuming complete $y$ coverage using 22(22, 44) bins and an integrated 
luminosity per bin of 2.5(5, 1.25)$fb^{-1}$, respectively. The corresponding 
bins widths are $\Delta y=$0.2(0.2, 0.1). The dotted curve 
corresponds to a 1 TeV NLC using 47 $\Delta y=0.2$ bins with 2.5 $fb^{-1}$/bin. 
`s' labels the SM prediction.}
\end{figure}

We begin by examining the energy, \ie, $y$ dependence of $A_{pol}$ 
for the two processes of interest; we consider the $W$ case first. For a 
500(1000) GeV collider, we see that only the range $1\leq y\leq 5.4(10.4)$ is 
kinematically accessible since the laser photon energy 
maximum is $\simeq 0.84E_e$. Since we are interested in bounds on the 
anomalous couplings, we assume that the SM is valid and generate a set 
of binned $A_{pol}$ data samples via Monte Carlo taking 
only the statistical errors into account. We further assume that 
the electrons are 
90$\%$ left-handed polarized as right-handed electrons do not interact 
through the $W$ charged current couplings. Our bin width will be assumed to be 
$\Delta y=$0.1 or 0.2. 
We then fit the resulting distribution to 
the $\Delta \kappa$- and $\lambda$-dependent functional form of $A_{pol}(y)$ 
and subsequently 
extract the 95$\%$ CL allowed ranges for the anomalous couplings. The results 
of this procedure are shown in Fig. 1, where we see that reasonable 
constraints are obtained although only a single observable has been used in 
the fit. 

\vspace*{-0.5cm}
\nn
\begin{figure}[htbp]
\centerline{
\psfig{figure=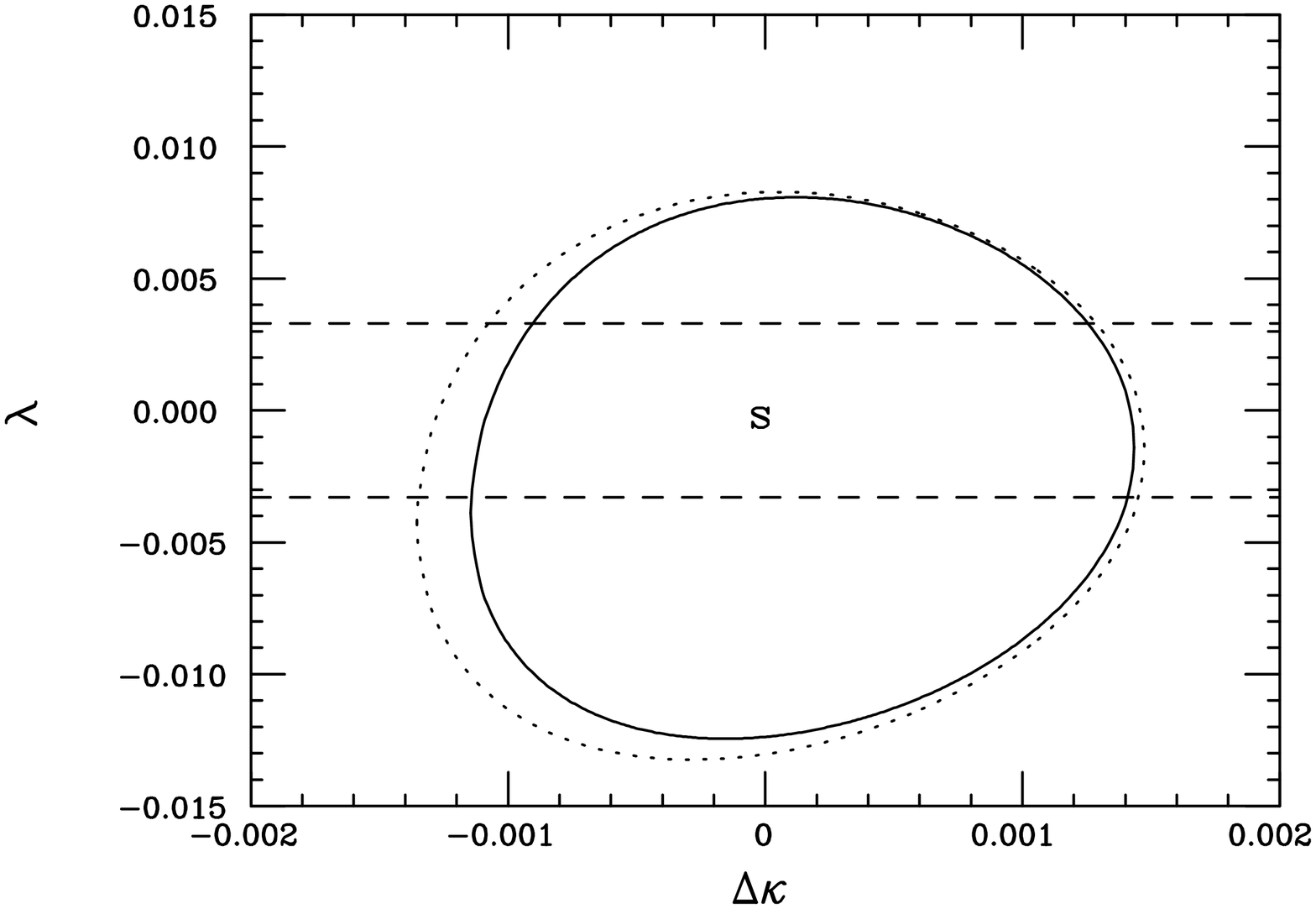,height=5.5cm,width=6.5cm,angle=-90}
\hspace*{-5mm}
\psfig{figure=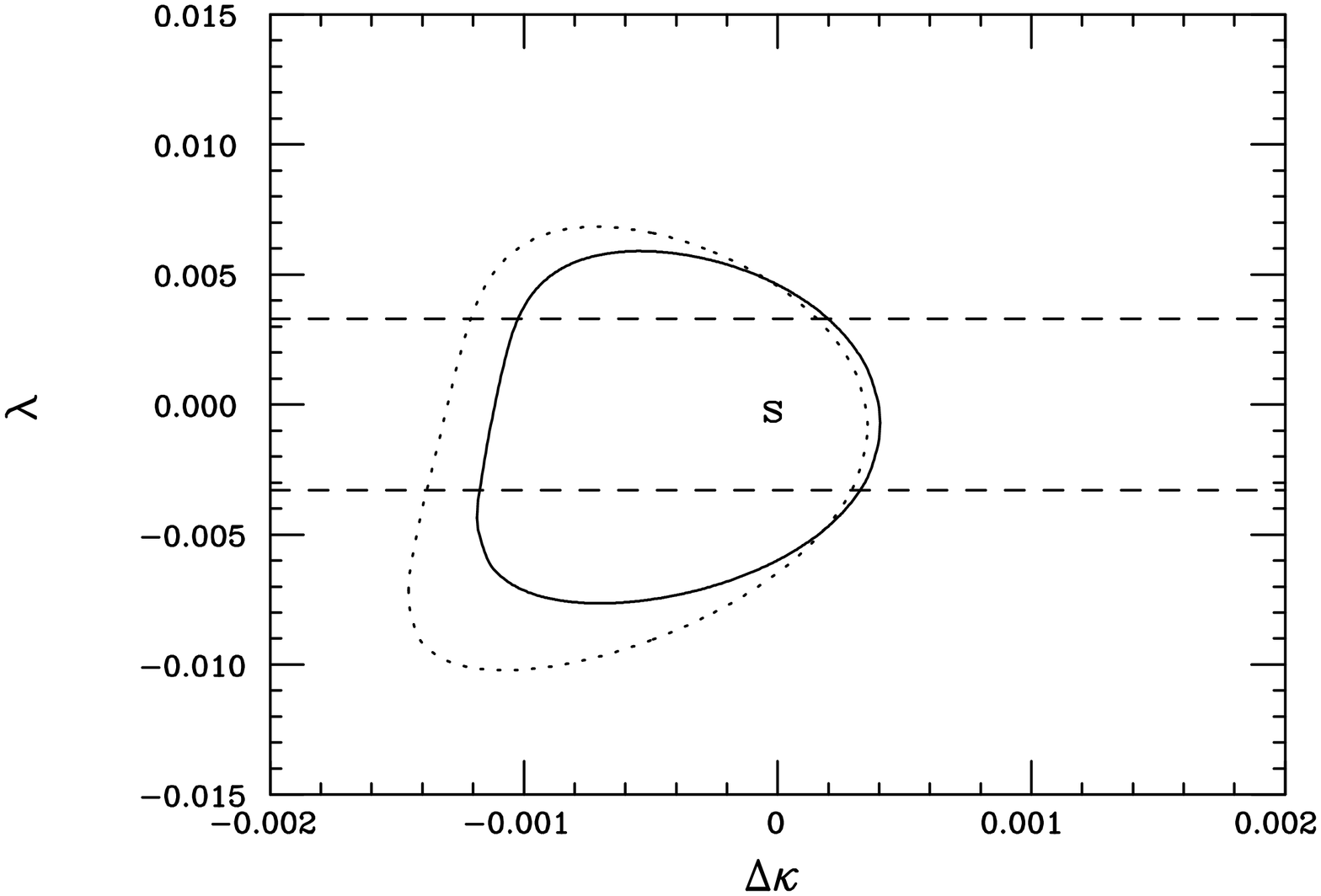,height=5.5cm,width=6.5cm,angle=-90}}
\vspace*{-1cm}
\caption{Same as the previous figure, but now for a (0.5)1 TeV NLC on the 
left(right) and combined with data 
on the total cross section and angular distribution in a simultaneous fit. The 
dotted(solid) curve uses the polarization 
asymmetry and total cross section(all) data. Only 
statistical errors are included. The dashed lines are the corresponding 
bounds from the LHC from the $pp\to W\gamma +X$ process with an integrated 
luminosity of 100 $fb^{-1}$.}
\end{figure}
\vspace*{0.4mm}
\vspace*{-0.5cm}
\nn
\begin{figure}[htbp]
\centerline{
\psfig{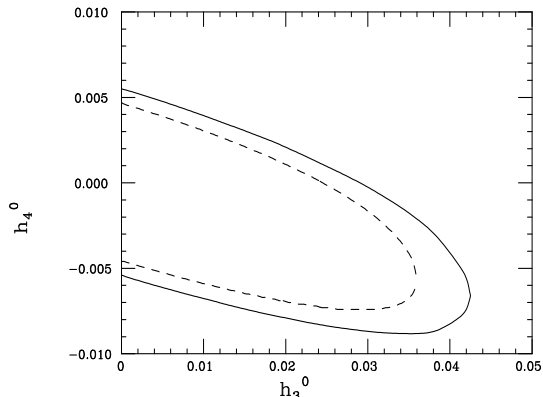}}
\vspace*{-1cm}
\caption{$95\%$CL allowed region for the anomalous coupling parameters 
$h_3^0$ and $h_4^0$ from a combined fit to the energy dependencies of the total 
cross section and polarization asymmetry at a 500 GeV NLC assuming $P=90\%$ 
and an integrated luminosity of $3(6)fb^{-1}$/bin corresponding to the solid
(dashed) curve. 18 bins of width $\Delta y$=0.2 were chosen to cover the $y$ 
range $1\leq y \leq 4.6$. The corresponding bounds for negative values of 
$h_3^0$ are obtainable by remembering the invariance of the polarization 
dependent cross section under the reflection $h_{3,4}^0\to -h_{3,4}^0$.}
\end{figure}
\vspace*{0.4mm}

Clearly, to obtain stronger limits we need to make a combined fit with other 
observables, such as the energy dependence of the total cross section, the 
$W$ angular distribution, or the net $W$ polarization. As an example we 
show in Fig. 2 from the results of our Monte Carlo study that the size of 
the $95\%$ CL allowed region shrinks drastically in the both the 0.5 and 
1 TeV cases when the $W$ angular distribution and energy-dependent total 
cross section data are included in a simultaneous fit together 
with the polarization asymmetry. Note that the constraints obtained by this 
analysis are comparable to those obtainable at the LHC~{\cite {rev}}.

In the $Z$ case we will follow a similar approach but we 
will simultaneously fit both the energy dependence of $A_{pol}$ as well as 
that of 
the total cross section. (Later, we will also include the $Z$ boson's angular 
distribution 
into the fit.) In this $Z$ analysis we make a $10^{\circ}$ angular cut on the 
outgoing electron and keep a finite form factor scale, $\Lambda=1.5$ TeV, so 
that we may more readily compare with other existing analyses. (The angular 
cut also gives us a finite cross section in the massless electron limit; 
this cut is not required in the case of the $W$ production process.) We again  
assume that $P=90\%$ so that data taking for this analysis can take place 
simultaneously with that for the $W$. The accessible $y$ range is
$1\leq y \leq 4.6$ for a 500 GeV collider; Fig.3 shows our results 
for this case.

Even these anomalous coupling bounds can be significantly 
improved by including the $Z$ boson angular information in 
the fit. The 
result of this procedure is shown in Fig.4 together with 
the anticipated result from the LHC using the $Z\gamma$ production mode. Note 
that the additional angular distribution data has reduced the size of the 
$95\%$ CL allowed region by almost a factor of two. 
Clearly both machines are complementary in their abilities to probe small 
values of the $\gamma ZZ$ anomalous couplings. As in the $W$ case, if 
the NLC and LHC results were 
to be combined, an exceptionally small allowed region would remain. 
The NLC results themselves may be further improved by considering 
measurements of the polarization of the final state $Z$ as well as 
by an examination of, \eg, the complementary $e^+e^- \to Z\gamma$ process. 

\vspace*{-0.5cm}
\nn
\begin{figure}[htbp]
\centerline{
\psfig{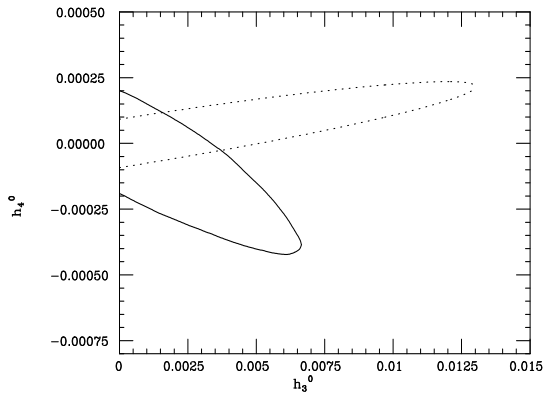}}
\vspace*{-1cm}
\caption{The solid curve is the same as dash-dotted curve 
in Fig. 3, but now at a center of mass energy of 1 TeV and including in the fit 
the $Z$ boson angular distribution obtained from the highest 
10 bins in energy. The 
corresponding result for the 14 TeV LHC with 100$fb^{-1}$ of integrated 
luminosity from the process $pp\to Z\gamma+X$ is shown as the dotted curve.}
\end{figure}
\vspace*{0.4mm}

The collision of polarized electron and photon beams at the NLC offers an 
exciting opportunity to probe for anomalous gauge couplings of both the $W$ 
and the $Z$ through the use 
of the polarization asymmetry. In the case of $\gamma e \to W\nu$ we can 
cleanly isolate the $\gamma WW$ vertex in a model independent fashion. When 
combined with other observables, extraordinary sensitivities to such 
couplings for $W$'s are achievable at the NLC in the $\gamma e$ mode. These are 
found to be quite complementary to those obtainable in $e^+e^-$ collisions as 
well as at the LHC. In the case of the 
$\gamma ZZ$ anomalous couplings, we found constraints comparable to those 
which can be obtained at the LHC.

\section*{Acknowledgments}

The author would like to thank J.L. Hewett, S.J. Brodsky and I. Schmidt
for discussions related to this work. 

\section*{References}

%
%%%%%%%%%%%%%%%%%%--- References
%%%%%%%%%%%%%%%%%%%%%%%%%%%%%%%%%%%%%%%%%%%%%%%%%%%%%%%
\def\MPL #1 #2 #3 {Mod.~Phys.~Lett.~{\bf#1},\ #2 (#3)}
\def\NPB #1 #2 #3 {Nucl.~Phys.~{\bf#1},\ #2 (#3)}
\def\PLB #1 #2 #3 {Phys.~Lett.~{\bf#1},\ #2 (#3)}
\def\PR #1 #2 #3 {Phys.~Rep.~{\bf#1},\ #2 (#3)}
\def\PRD #1 #2 #3 {Phys.~Rev.~{\bf#1},\ #2 (#3)}
\def\PRL #1 #2 #3 {Phys.~Rev.~Lett.~{\bf#1},\ #2 (#3)}
\def\RMP #1 #2 #3 {Rev.~Mod.~Phys.~{\bf#1},\ #2 (#3)}
\def\ZP #1 #2 #3 {Z.~Phys.~{\bf#1},\ #2 (#3)}
\def\IJMP #1 #2 #3 {Int.~J.~Mod.~Phys.~{\bf#1},\ #2 (#3)}

\end{document}